\documentclass{ifacconf}

\usepackage{graphicx}      
\usepackage{natbib}        
\usepackage{subcaption}
\usepackage{tikz}
\usepackage{amsmath,amssymb,amsfonts}
\usepackage{float}
\usepackage{url}
\usepackage[T1]{fontenc}
\usepackage{lmodern}
\usepackage{cmap}
\usetikzlibrary{shapes, arrows.meta, decorations.pathreplacing, positioning, calc, shapes.multipart, fit, backgrounds, patterns, intersections, matrix, shadows.blur,overlay-beamer-styles}
\definecolor{CustomBlue}{RGB}{0, 86, 164}
\colorlet{CustomBlueLight}{CustomBlue!20!white}
\colorlet{GreenLight}{green!40!white}
\colorlet{RedLight}{red!20!white}
\colorlet{GreyLight}{gray!20!white}
\definecolor{Grey}{RGB}{150,150,150}
\tikzset{
	myarrow/.style={-Stealth, draw=black}
}

\begin{document}
\begin{frontmatter}

\title{Physics-Informed Neural Network for Modeling the Dynamic Behavior of Grid-Forming Converters \thanksref{footnoteinfo}} 

\thanks[footnoteinfo]{Simulations were performed with computing resources granted by RWTH Aachen University under project \textbf{thes2128}.}
\thanks[footnoteinfo]{© 2026 the authors. This work has been accepted by IFAC for publication under a Creative Commons license CC-BY-NC-ND.}

\author[First]{Hussein Jaffal} 
\author[First]{Arianna Fois}
\author[First]{Sarra Bouchkati} 
\author[First]{Amirali Mahjoob}
\author[First]{Andreas Ulbig}

\address[First]{Institute of High Voltage Equipment and Grids, Digitalization and Energy Economics, RWTH Aachen University, Aachen, Germany (e-mail: h.jaffal@iaew.rwth-aachen.de).}


\begin{abstract}                
This paper investigates physics-informed neural networks for modeling the full dynamic behavior of droop-controlled grid-forming converters. The approach is trained on synthetic data generated via numerical solvers and benchmarked against both traditional integration methods and a vanilla neural network. Results show higher predictive accuracy than the vanilla network using the same training data and substantially reduced runtime compared with numerical solvers.

\end{abstract}


\end{frontmatter}

\section{Introduction}
\subsection{Background and Motivation}
The increasing integration of renewable energy sources (RES) is essential for achieving net-zero emission targets by 2050. Unlike conventional synchronous generators, RES are typically connected to the grid through power electronic converters. The rising penetration of such inverter-based resources (IBRs) introduces fast and nonlinear dynamics, creating new challenges for grid stability \cite{erdiwansyah2021critical}. Power system operators are therefore mandated to perform dynamic stability assessments (DSA), as outlined in \cite{vision2022power}. The dynamic behavior of power system components, including IBRs, can be represented by ordinary differential equations (ODEs). Although these equations yield detailed and broadly applicable models, solving them efficiently, i.e., balancing computational cost and accuracy, remains difficult, especially as system size and IBR penetration grow. Traditional Root Mean Square (RMS) models are widely used for their simplicity and speed, but their strong assumptions make them unsuitable for capturing the fast, complex dynamics associated with converters \cite{misyris2021grid}. As a result, Electromagnetic Transient (EMT) models are increasingly adopted because they provide high-fidelity representations of converter behavior \cite{winkens2021impact}. Their limitation, however, is the requirement for very small time steps to maintain accuracy. In practice, EMT simulations rely on numerical solvers such as Runge–Kutta (RK) methods, which, despite their stability and modularity, become computationally expensive for fast dynamics and large-scale systems. Even with recent improvements, solver performance remains constrained by fundamental numerical limits \cite{liu2019solving}. To address these challenges, machine-learning (ML)-based methods have been proposed as an alternative for modeling converter dynamics with reduced computational time, provided sufficient data are available.
\subsection{Related Works}
Data-driven methods and ML techniques have recently gained significant attention in power system modeling \cite{duchesne2020recent}. Several works, such as \cite{stiasny2021transient}, have implemented vanilla neural networks (VNNs) to accelerate dynamic simulations by learning complex system behavior. However, their performance depends heavily on large and diverse datasets to ensure generalization \cite{ahmad2024interpretable}, and purely data-driven models may produce results that are not physically consistent. Another approach is Neural Ordinary Differential Equations (Neural ODEs) \cite{chen2018neural}, which learn the underlying ODE function rather than its solution, preserving the structure of classical integration schemes. Neural ODEs have been applied to multi-component dynamic simulations \cite{xiao2022feasibility, bossart2025acceleration}, but have not yet shown substantial speed advantages over conventional solvers.

Another emerging approach proposed as a promising alternative to purely data-driven methods is the Physics-Informed Neural Network (PINN). By embedding physical dynamic equations into the learning process, PINNs reduce the need for large labeled datasets while ensuring physically consistent predictions and mitigating overfitting. Prior works \cite{stiasny2021learning,stiasny2023physics, misyris2020physics, ellinas2025physics} applied PINNs to generator dynamics, demonstrating their ability to accelerate time-domain simulations and accurately capture system behavior, but these studies were limited to synchronous generators. More recently, \cite{nellikkath2024physics} used PINNs to model grid-following converters (GFL), focusing on the phase-locked loop (PLL), and successfully predicted only two GFL-related states. Another study \cite{jagadeesan2024enhanced} attempted to model all GFL states with PINNs, but the training process failed to converge. Overall, existing research remains limited: prior studies either focus solely on generators, consider only a small subset of grid-following converter states, or do not achieve successful training when modeling the full converter dynamics.
\subsection{Contribution and Structure}
In this paper, we present an approach to train a PINN for capturing the full dynamics of a droop-controlled grid-forming converter (GFM), based on the EMT model in~\cite{winkenssest}. To the best of our knowledge, this is the first work to propose a PINN framework capable of learning the full state dynamics of a grid-forming converter. The method advances the application of PINNs in power system modeling and offers potential speedups for time-domain simulations. We summarize our contributions as follows:

\begin{enumerate}
    \item We extend the application of PINN to grid-forming converters, demonstrating that PINNs can effectively learn their fast and nonlinear dynamics.
    \item We model the complete state dynamics of grid-forming converters, incorporating all relevant control loops and the grid interface to provide a high-fidelity representation of the system.
\end{enumerate}


The remainder of the paper is structured as follows: Section~\ref{sec:PF} provides background on ODEs, GFM, and PINNs. Section~\ref{sec: Metho} outlines the proposed modeling approach. Section~\ref{sec: Results} presents and evaluates the results, and Section~\ref{sec: Conc} concludes with future work.

\section{Problem Formulation}
\label{sec:PF}
In this section, we provide the necessary background of ODEs, the problem formulation of the dynamics of the GFM, as well as introduction to PINNs. 
\subsection{Ordinary Differential Equations}
\label{ODE}
A dynamic system governed by an ordinary differential equation describes the evolution of the system state \( x \in \mathbb{R}^n \) over time \( t \in \mathbb{R}\), where the state derivative is given by a function that depends on the state itself, time \(t\) and the input \( u \). Formally, this is expressed as:
\begin{equation}
    \frac{d\textbf{x}}{dt} = \textbf{f}(\textbf{x}, \textbf{t}, \textbf{u)}, \; \; \space  f : \mathbb{R}^n \times \mathbb{R} \times
\mathbb{R}^m \rightarrow \mathbb{R}^n \
    \label{eq:ODE}
\end{equation}

The main task, therefore, is to learn a function \( f \) that accurately approximates the solution of the underlying ODE, that is, learns the dynamics of the system over time. Therefore, the central idea is to approximate the solution \( x(t) \) using a PINN. This can be formulated as:

\begin{equation}
    x(t) \approx \hat{x}(t) = \text{PINN}(x_0, t, u)
\end{equation}

where $\hat{x}(t)$ is the approximate solution.

\subsection{Grid-forming Converter Architecture}
\label{subsec:model}
  
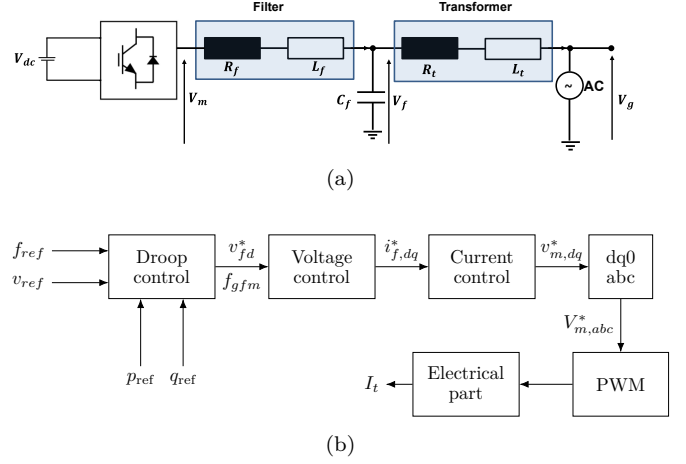
\begin{figure}[htbp]
    \centering
    
    \begin{subfigure}[b]{\linewidth}
        \centering
        \includegraphics[width=\linewidth]{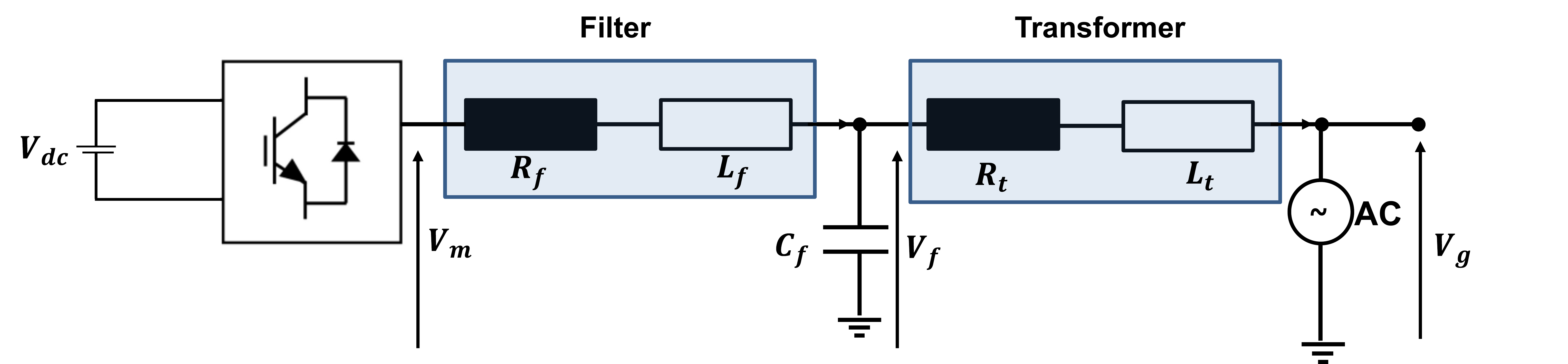}
        \caption{}
        \label{fig:fig1a}
    \end{subfigure}
    
    \vspace{0.5cm}
    
    \begin{subfigure}[b]{\linewidth}
        \centering
        \resizebox{\linewidth}{!}{%
        \begin{tikzpicture}[%
          block/.style={draw, minimum width=2cm, minimum height=1.2cm, align=center},
          smallblock/.style={draw, minimum width=1.2cm, minimum height=1.2cm, align=center},
          >=Latex
        ]
        
        \node[block] (droop) {Droop\\control};
        \node[block, right=1cm of droop] (voltage) {Voltage\\control};
        \node[block, right=1cm of voltage] (current) {Current\\control};
        \node[smallblock, right=1cm of current] (dq0abc) {dq0\\abc};
        \node[block, below=1cm of dq0abc, minimum width=1.8cm] (pwm) {PWM};
        \node[block, left=1cm of pwm] (electrical) {Electrical\\part};

        \draw[->] (droop.east) -- (voltage.west) node[pos=0.5,above]{$v_{fd}^*$} node[pos=0.5,below]{$f_{gfm}$};
        \draw[->] (voltage) -- (current) node[pos=0.5, above]{$i_{f,dq}^*$}; 
        \draw[->] (current) -- (dq0abc) node[pos=0.5, above]{$v_{m,dq}^*$};
        \draw[->] (dq0abc.south) -- (pwm.north) node[pos=0.5, left] {$V_{m,abc}^*$};
        \draw[->] (pwm.west) -- (electrical.east);

        \node[xshift=-0.4cm, yshift=-1.5cm] at (droop.south) (p) {$p_{\text{ref}}$};
        \node[xshift=0.4cm, yshift = -1.5cm] at (droop.south) (q) {$q_{\text{ref}}$};
        \node[xshift=-1.5cm, yshift=0.3cm] at (droop.west) (f) {$f_{ref}$};
        \node[xshift=-1.5cm, yshift=-0.3cm] at (droop.west) (v) {$v_{ref}$};
        \draw[->] (p.north) -- ([xshift =-0.4cm] droop.south);
        \draw[->] (q.north) -- ([xshift = 0.4cm] droop.south);
        \draw[->] (f.east) -- ([yshift=0.3cm] droop.west);
        \draw[->] (v.east) -- ([yshift=-0.3cm] droop.west);

        \node[left=0.5cm of electrical] (out) {$I_t$};
        \draw[->] (electrical.west) -- (out.east);

        \end{tikzpicture}
        } 
        \caption{}
        \label{fig:fig1b}
    \end{subfigure}
    
    \caption{Equivalent circuit (a) and general control structure (b) of the GFM}
    \label{fig:mainfigure}
\end{figure}

The general equivalent circuit of the converter and the internal control architectures used as the foundation for deriving the non-linear dynamic equations are shown in Figures \ref{fig:fig1a} and \ref{fig:fig1b}, respectively.

A detailed overview of each block in Figure \ref{fig:fig1b}, as well as the electrical part shown in Figure \ref{fig:fig1a}, is presented in Figures~\ref{fig:mainfigure}, \ref{fig:volin}, and \ref{fig:currin}. Specifically, the following subsections describe and derive the dynamic equations associated with each block in Figure \ref{fig:fig1b}, which serve as the foundation for training the physics-informed neural network.

For the remainder of this paper, $V_m$ denotes the converter-side voltage, $V_f$ the filter capacitor voltage, and $V_g$ the PCC voltage. The current $I_f$ flows through the filter impedance, defined by $R_f$ and $L_f$, with $C_f$ as the filter capacitance. The subscripts \textit{d} and \textit{q} refer to the direct and quadrature components in the rotating $dq$ frame. The grid interface is modeled using $R_t$ and $L_t$. Component specifications and tuning parameters are provided in \cite{winkenssest}. All parameters with superscript $^{*}$ denote setpoints, while the subscripts $d$ and $q$ indicate the components in the $dq$-reference frame.

\subsubsection{Electrical Part}
The electrical part represents the converter’s interface with the grid, including the filter, the grid connection, and the external grid. Based on Figure~\ref{fig:fig1a}, the dynamic equations can be derived by applying Kirchhoff’s Voltage Law (KVL) and Kirchhoff’s Current Law (KCL):
\begin{align}
    V_m - V_f &= R_f \cdot I_f + L_f \cdot \frac{dI_f}{dt} \\
    I_f - I_t &= C_f \cdot \frac{dV_f}{dt} \\
    V_f - V_g &= R_t \cdot I_t + L_t \cdot \frac{dI_t}{dt}
    \label{General_eq}
\end{align}

Applying the dq$-$transformation and normalizing the equations to the per-unit system, these set of equations can be expressed in the dq$-$reference frame rotating at an angular speed $\omega_{gfm}$ with a phase angle of $\theta_{gfm}$. 


\subsubsection{Droop Control}

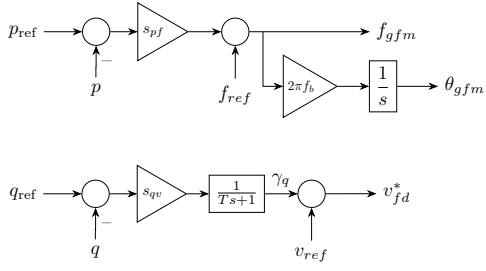
\begin{figure}[htbp]
    \centering
    \begin{subfigure}[b]{\linewidth}
        \centering
        \resizebox{0.75\linewidth}{!}{
        \begin{tikzpicture}[auto, node distance=1.8cm and 1.5cm, >=Latex, scale=0.5]
            \node (pset) at (0,0) {\(p_{\text{ref}}\)};
            \node[coordinate, right=-0.8cm of pset] (sumP) {};
            \node[circle, draw, minimum size=5mm, right=of sumP] (sumPcircle) {};
            \node[isosceles triangle, draw, shape border rotate=360, isosceles triangle apex angle=60, minimum height=0.3cm, right=0.5cm of sumPcircle, scale =0.8] (kp) {\(s_{pf}\)};
            \node[right=0.6cm of kp] (sumF) [circle, draw, minimum size=5mm] {};
            \node[below=0.2cm of sumF] {};
            \node[right= 2.2cm of sumF] (fout) {\(f_{gfm}\)};
            \node[below=0.6cm of pset, xshift=1.335cm] (p) {\(p\)};
            \node[below=0.6cm of sumF] (f) {\(f_{ref}\)};
            \node[below=0.15cm of sumPcircle,xshift = 0.2cm] {–};
            \node[isosceles triangle, draw, shape border rotate=360, isosceles triangle apex angle=60, minimum height=0.3cm, below of= fout, node distance=1cm, scale =0.7, xshift=-2.5cm] (wb) {\(2 \pi f_b\)};
            \node (int) [draw, rectangle, minimum width=0.5cm, minimum height=0.5cm, right of=wb, node distance=1.5cm] {$\displaystyle \frac{1}{s}$};
            \node (theta) [right of=int, node distance=1.5cm]{$\theta_{gfm}$};
            \draw[myarrow] (p) -- ++(0,0.85) -| (sumPcircle) ;
            \draw[myarrow] (f) -- ++(0,0.85) -| (sumF);
            \draw[myarrow] (wb.east) -- (int.west);
            \draw[myarrow] (int.east) -- (theta.west);
            \draw[myarrow] ($(sumF.east)+(0.5,0)$) |- (wb.west);
            \draw[myarrow] (pset) -- (sumPcircle);
            \draw[myarrow] (sumPcircle) -- node {} (kp);
            \draw[myarrow] (kp) -- (sumF);
            \draw[myarrow] (sumF) -- (fout);

            \node[below=2.5cm of pset] (qset) {\(q_{\text{ref}}\)};
            \node[coordinate, right=-0.8cm of qset] (sumQ) {};
            \node[circle, draw, minimum size=5mm, right=of sumQ] (sumQcircle) {};
            \node[below=0.15cm of sumQcircle,xshift = 0.2cm] {–};
            \node[isosceles triangle, draw, shape border rotate=360, isosceles triangle apex angle=60, minimum height=0.3cm, right=0.5cm of sumQcircle, scale =0.8] (kq) {\(s_{qv}\)};
            
            \node[draw, rectangle, right=0.4cm of kq, minimum width=10mm, minimum height=6mm] (transfer) {\(\frac{1}{Ts + 1}\)};
            \node[right=0.6cm of transfer] (sumV) [circle, draw, minimum size=5mm] {};
            \node[below=0.6cm of qset, xshift = 1.3255cm] (q) {\(q\)};
            \draw[myarrow] (q) -- ++(0,0.85) -| (sumQcircle);
            \node[below=0.6cm of sumV] (V) {\(v_{ref}\)};
            \node[below=2.35cm of fout] (Vst) {$v_{fd}^*$};
            \draw[myarrow] (qset) -- (sumQcircle);
            \draw[myarrow] (sumQcircle) -- (kq);
            \draw[myarrow] (kq) -- (transfer);
            \draw[myarrow] (transfer) -- (sumV) node[pos=0.5,above]{$\gamma_q$};
            \draw[myarrow] (sumV) -- (Vst);
            \draw[myarrow] (V.north) -- (sumV.south);
        \end{tikzpicture} }
        \label{fig:outp}
    \end{subfigure}
    \caption{Outer active and reactive control loops of the GFM}
    \label{fig:droop}
\end{figure}

Figure~\ref{fig:droop} shows the active and reactive control loops, representing the outer control of the GFM. 

In the active power loop, the reference power $p_{\text{ref}}$ is compared with the measured power $p$ to determine the converter’s frequency and phase dynamics. These dynamics are governed by the droop gain $s_{pf}$, the base frequency $f_b$, and the reference frequency $f_{\text{ref}}$, and are described as follows:
\begin{align}
    f_{\text{gfm}} &= s_{pf} \cdot (p_{\text{ref}} - p) + f_{\text{ref}} \label{eq:fgfm} \\
    \dot{\theta}_{\text{gfm}} &= 2\pi f_b \cdot f_{\text{gfm}} \label{eq:theta_gfm} 
\end{align}
The reactive power loop has a similar structure, where the reference $q_{\text{ref}}$ is compared with the measured $q$. The error is processed through a droop gain $s_{qv}$ and a PT1 filter with cut-off frequency $\omega_q$, leading to the following dynamics:

\begin{align}
    \dot{\gamma}_q &= -\omega_q \gamma_q + \omega_q s_{qv}(q_{\text{ref}} - q) \label{eq:gamma_q} \\
    v_{fd}^* &= v_{\text{ref}} + \gamma_q \label{eq:vfd_star} 
\end{align}
\subsubsection{Inner Voltage Control Loop}
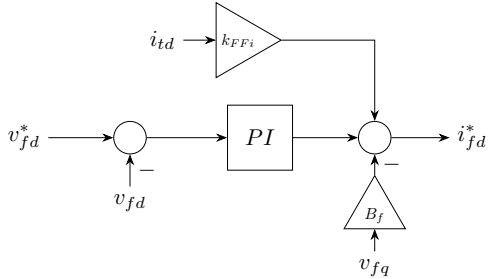
\begin{figure}[htbp]
    \centering
    \resizebox{0.75\linewidth}{!}{
    \begin{tikzpicture}
		\node (vcapds) at (0,0) {$v_{fd}^*$};
		\node[circle, draw, minimum size=5mm, right=of vcapds, node distance=1cm] (c1) {};
		\node (PI1) [draw, rectangle, minimum width=1cm, minimum height = 1cm, right of=c1, node distance = 2cm] {$PI$};
		\node[circle, draw, minimum size=5mm, right=of PI1, node distance=1cm] (c2) {};
        \node (itd) [above of=PI1, node distance=1.5cm, xshift=-1.5cm]{$i_{td}$};
        \node[isosceles triangle, draw, shape border rotate=360, isosceles triangle apex angle=60, minimum height=0.3cm, right=0.5cm of itd, scale =0.7] (kp) {$k_{FFi}$};
        \node (ifd) [right of=c2, node distance=1.5cm]{$i_{fd}^*$};
        \node (vfq) [below of=c2, node distance=2cm]{$v_{fq}$};
        \node (vfd) [below of=c1, node distance=1cm]{$v_{fd}$};
        \node[isosceles triangle, draw, shape border rotate=90, isosceles triangle apex angle=60, minimum height=0.3cm, above of=vfq, scale =0.7, node distance=0.8cm] (bf) {$B_f$};
        \draw[myarrow] (vcapds.east) -- (c1.west);
        \draw[myarrow] (vfd.north) -- (c1.south) node [pos=0.5, right]{$-$};
        \draw[myarrow] (itd.east) -- (kp.west);
        \draw[myarrow] (kp.east) -| (c2.north);
        \draw[myarrow] (c1.east) -- (PI1.west);
        \draw[myarrow] (PI1.east) -- (c2.west);
        \draw[myarrow] (c2.east) -- (ifd.west);
        \draw[myarrow] (vfq.north) -- (bf.south);
        \draw[myarrow] (bf.north) -- (c2.south) node[pos=0.5, right] {$-$};
	\end{tikzpicture}}
    \caption{Inner voltage control loop of the GFM}
    \label{fig:volin}
\end{figure}

As depicted in Figure~\ref{fig:volin}, the current $i_{fd}$ is regulated by controlling the voltage $v_{fd}$, which is handled by the inner voltage control loop. 

The corresponding dynamics are described by the following equation. An auxiliary state variable $\xi_d$ is introduced to represent the integral action of the PI controller in the voltage control loop. The parameters $k_{pv}$ and $k_{iv}$ denote the proportional and integral gains, respectively. Although only the $d$-axis component is shown in Figure~\ref{fig:volin}, the same structure is applied to the $q$-axis:
\begin{align}
    \dot{\xi_d} &= v_{fd}^* - v_{fd}  \label{eq:xi_d}
\end{align}
\begin{align}
    i_{fd}^* &= k_{pv} \cdot (v_{fd}^* - v_{fd}) + k_{iv} \cdot \xi_d + k_{\text{FF}i} \cdot i_{td} - \omega_{\text{gfm}} c_f \cdot v_{fq} \label{eq:ifd_star}
\end{align}

\subsubsection{Inner Current Control Loop}
The output reference current $i_{fd}^*$ is passed to the inner current control loop, as shown in Figure~\ref{fig:currin}. An auxiliary state variable $\sigma_d$ is introduced to represent the integral action of the controller. Although only the $d$-axis component is illustrated, the same structure applies to the $q$-axis. The corresponding dynamics are given as follows:

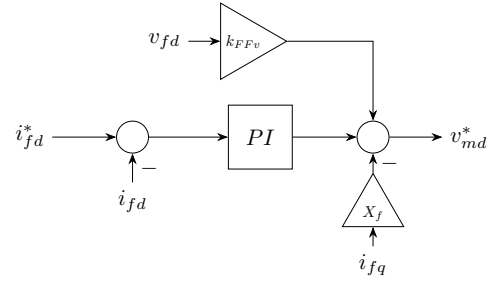
\begin{figure}[htbp]
    \centering
    \resizebox{0.75\linewidth}{!}{
    \begin{tikzpicture}
		\node (vcapds) at (0,0) {$i_{fd}^*$};
		\node[circle, draw, minimum size=5mm, right=of vcapds, node distance=1cm] (c1) {};
		\node (PI1) [draw, rectangle, minimum width=1cm, minimum height = 1cm, right of=c1, node distance = 2cm] {$PI$};
		\node[circle, draw, minimum size=5mm, right=of PI1, node distance=1cm] (c2) {};
        \node (itd) [above of=PI1, node distance=1.5cm, xshift=-1.5cm]{$v_{fd}$};
        \node[isosceles triangle, draw, shape border rotate=360, isosceles triangle apex angle=60, minimum height=0.3cm, right=0.5cm of itd, scale =0.7] (kp) {$k_{FFv}$};
        \node (ifd) [right of=c2, node distance=1.5cm]{$v_{md}^*$};
        \node (vfq) [below of=c2, node distance=2cm]{$i_{fq}$};
        \node (vfd) [below of=c1, node distance=1cm]{$i_{fd}$};
        \node[isosceles triangle, draw, shape border rotate=90, isosceles triangle apex angle=60, minimum height=0.3cm, above of=vfq, scale =0.7, node distance=0.8cm] (bf) {$X_f$};
        \draw[myarrow] (vcapds.east) -- (c1.west);
        \draw[myarrow] (vfd.north) -- (c1.south) node [pos=0.5, right]{$-$};
        \draw[myarrow] (itd.east) -- (kp.west);
        \draw[myarrow] (kp.east) -| (c2.north);
        \draw[myarrow] (c1.east) -- (PI1.west);
        \draw[myarrow] (PI1.east) -- (c2.west);
        \draw[myarrow] (c2.east) -- (ifd.west);
        \draw[myarrow] (vfq.north) -- (bf.south);
        \draw[myarrow] (bf.north) -- (c2.south) node[pos=0.5, right] {$-$};
	\end{tikzpicture} }
    \caption{Inner current control loop of the GFM}
    \label{fig:currin}
\end{figure}
\begin{align}
    \dot{\sigma}_d &= i_{fd}^* - i_{fd} 
    \label{eq:sigma_d}
\end{align}

\begin{align}
    v_{md}^* &= k_{p,c}(i_{fd}^* - i_{fd}) + k_{i,c}\,\sigma_d + k_{\text{FF}v} v_{fd} \notag \\
              &\quad - \omega_{\text{gfm}} l_f\, i_{fq}
    \label{eq:vmd_star}
\end{align}

The final system state vector $x_{gfm}$ and the input vector $u_{gfm}$ can be summarized as follows:
\begin{equation}
x_{gfm} = 
\begin{bmatrix}
v_{fd} & v_{fq} & \xi_d & \xi_q & i_{fd} & i_{fq} & \sigma_d & \sigma_q & i_{td} & i_{tq} \\
\gamma_q & \theta_{gfm} & \theta_{grid}
\end{bmatrix}^{\top}
\label{eq: states}
\end{equation}
\begin{equation}
u_{gfm} = 
\begin{bmatrix}
p_{ref} & q_{ref} & v_{ref} & f_{ref}
\end{bmatrix}^{\top}
\label{eq: inputs}
\end{equation}

\subsection{Physics-Informed Neural Network}
\begin{figure}[htbp]
    \centering
    \resizebox{0.7\linewidth}{!}{
		\begin{tikzpicture}
			\node (input1)[draw, circle, fill=CustomBlue!20!white] at (0,0){$t$};
			\node (input2)[below of= input1, node distance =1cm, draw, circle, fill=CustomBlue!20!white]{$x_0$};
			\node (input3)[below of= input2, node distance =1cm, draw, circle, fill=CustomBlue!20!white]{$u$};
			\node (sigma11)[right of= input1, node distance =1.5cm, draw, circle, fill=Grey!20!white]{$\sigma$};
			\node (sigma12)[above of= sigma11, node distance =1.2cm, draw, circle, fill=Grey!20!white]{$\sigma$};
			\node (sigma13)[below of= sigma11, node distance =1.2cm, draw, circle, fill=Grey!20!white]{$\sigma$};
			\node (dot4)[below of=sigma13, node distance = 1cm] {$.$};
			\node (dot5)[below of=dot4, node distance = 0.25cm] {$.$};
			\node (dot6)[below of=dot5, node distance = 0.25cm] {$.$};
			\node (sigma14)[below of= dot6, node distance =1cm, draw, circle, fill=Grey!20!white]{$\sigma$};
			\node (sigma21)[right of= sigma11, node distance =1.5cm, draw, circle, fill=Grey!20!white]{$\sigma$};
			\node (sigma22)[above of= sigma21, node distance =1.2cm, draw, circle, fill=Grey!20!white]{$\sigma$};
			\node (sigma23)[below of= sigma21, node distance =1.2cm, draw, circle, fill=Grey!20!white]{$\sigma$};
			\node (dot7)[below of=sigma23, node distance = 1cm] {$.$};
			\node (dot8)[below of=dot7, node distance = 0.25cm] {$.$};
			\node (dot9)[below of=dot8, node distance = 0.25cm] {$.$};
			\node (sigma24)[below of= dot9, node distance =1cm, draw, circle, fill=Grey!20!white]{$\sigma$};
			\node (output1)[draw, circle, fill=red!20!white, right of=sigma23, node distance =1.5cm] {$\hat{x_t}$};
			\node (deriv)[draw, rectangle, minimum height=0.5cm, minimum width=0.5cm, right of=output1, node distance=3cm, fill=CustomBlue!20!white] {$\frac{d}{dt}$};
			
			\node (lossd)[draw, rectangle, fill=CustomBlue!20!white, below of=deriv, node distance =1.5cm] {$\mathcal{L_{PDL}}$};
			\node (loss)[draw, rectangle, fill=CustomBlue!20!white, left of=lossd, node distance =2cm] {$\mathcal{L_{NNL}}$};
			\node (losscol)[draw, rectangle, fill=CustomBlue!20!white, left of=lossd, node distance =1cm, yshift=0.8cm] {$\mathcal{L_{PCL}}$};
			\node (lossic)[draw, rectangle, fill=CustomBlue!20!white, left of=lossd, node distance =1cm, yshift=-2.5cm] {$\mathcal{L_{ICL}}$};
			\node (x0)[below of=lossd, node distance=2.5cm, xshift=0.3cm] {$x_0$};
			\node (update)[draw, rectangle, minimum height=0.5cm, minimum width=0.5cm, below of=sigma24, node distance=1cm, xshift=1.5cm, fill=CustomBlue!20!white] {$min_{w,b} \mathcal{L}$};
			\draw [myarrow] (input1.east) -- (sigma12.west);
			\draw [myarrow] (input1.east) -- (sigma11.west);
			\draw [myarrow] (input1.east) -- (sigma13.west);
			\draw [myarrow] (input1.east) -- (sigma14.west);
			\draw [myarrow] (input2.east) -- (sigma12.west);
			\draw [myarrow] (input2.east) -- (sigma11.west);
			\draw [myarrow] (input2.east) -- (sigma13.west);
			\draw [myarrow] (input2.east) -- (sigma14.west);
			\draw [myarrow] (sigma22.east) -- (output1.west);
			\draw [myarrow] (sigma21.east) -- (output1.west);
			\draw [myarrow] (sigma23.east) -- (output1.west);
			\draw [myarrow] (sigma24.east) -- (output1.west);
			\draw [myarrow] (sigma12.east) -- (sigma21.west);
			\draw [myarrow] (sigma12.east) -- (sigma22.west);
			\draw [myarrow] (sigma12.east) -- (sigma23.west);
			\draw [myarrow] (sigma12.east) -- (sigma24.west);
			\draw [myarrow] (sigma11.east) -- (sigma21.west);
			\draw [myarrow] (sigma11.east) -- (sigma22.west);
			\draw [myarrow] (sigma11.east) -- (sigma23.west);
			\draw [myarrow] (sigma11.east) -- (sigma24.west);
			\draw [myarrow] (sigma13.east) -- (sigma21.west);
			\draw [myarrow] (sigma13.east) -- (sigma22.west);
			\draw [myarrow] (sigma13.east) -- (sigma23.west);
			\draw [myarrow] (sigma13.east) -- (sigma24.west);
			\draw [myarrow] (sigma14.east) -- (sigma21.west);
			\draw [myarrow] (sigma14.east) -- (sigma22.west);
			\draw [myarrow] (sigma14.east) -- (sigma23.west);
			\draw [myarrow] (sigma14.east) -- (sigma24.west);
			\draw [myarrow] (input3.east) -- (sigma12.west);
			\draw [myarrow] (input3.east) -- (sigma11.west);
			\draw [myarrow] (input3.east) -- (sigma13.west);
			\draw [myarrow] (input3.east) -- (sigma14.west);
			\draw [myarrow] (update.west) -- ++(-1.6,0) -- ++(0,1) node [below] {$w_i , b_i$};
			\draw [myarrow] (deriv.south) -- (lossd.north);
			\draw [myarrow] (output1.east) -- (deriv.west);
			\draw [myarrow] (lossd.south) |- (update.east) ;
			\draw [myarrow] (output1.east) -- ++(0.5,0) -- ++(0,-1.22);
			\draw [myarrow] (loss.west) -| (update.north) ;
			\draw [myarrow] (x0.west) -- (lossic.east);
			\draw [myarrow] (deriv.south) |-(losscol.east);
			\draw [myarrow] (lossic.west) -| (update.south);
			\draw [myarrow] (losscol.south) |- (update.east);
		\end{tikzpicture}}
    \caption{Generalized architecture of the PINN}
    \label{fig:PINN}
\end{figure}
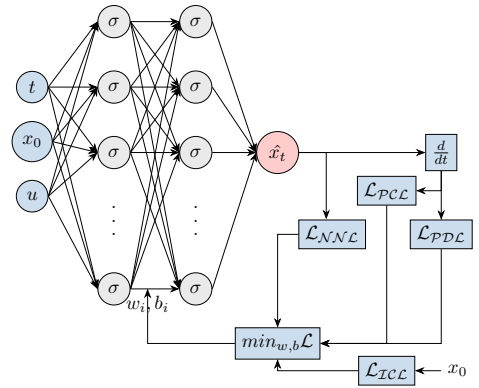

PINNs extend VNNs by embedding dynamic equations directly into the learning process. This incorporation improves generalization and greatly reduces the required training data. The general structure of a PINN is shown in Figure~\ref{fig:PINN}.

Based on the ODE structure in Section~\ref{ODE}, the time derivative of the network prediction is obtained through automatic differentiation. To enforce the system physics, three additional loss terms are added to the standard data-driven loss following~\cite{karampinis2025pinnstoolbox}. First, the standard neural network loss (NNL) is defined as:

\begin{equation}
    \mathcal{L}_{\text{NNL}} = \frac{1}{N_d} \sum_{i=1}^{N_d} \left| \hat{x}^{(i)} - x^{(i)} \right|^2
    \label{eq:LNN}
\end{equation}

The loss in \eqref{eq:LNN} corresponds to the standard data-driven loss used in the VNN framework, based on labeled input--output data. It minimizes the deviation between the network predictions and the ground truth, and is optimized with respect to the trainable parameters. The first additional loss is referred to as the \emph{physics data loss (PDL)} and is defined as:

\begin{equation}
    \mathcal{L}_{\text{PDL}} = \frac{1}{N_d} \sum_{i=1}^{N_d} \left| \frac{d\hat{x}^{(i)}}{dt} - f(t^{(i)}, x^{(i)}, u) \right|^2
    \label{eq:PLD}
\end{equation}

The loss in \eqref{eq:PLD} enforces consistency between the time derivative of the network prediction $\frac{d\hat{x}}{dt}$ and the system dynamics evaluated at the ground-truth states $x$. To enforce the initial conditions, the network output at time $t_0$ is constrained to match the known initial states through the \emph{initial condition loss (ICL)}, defined as:

\begin{equation}
    \mathcal{L}_{\text{ICL}} = \frac{1}{N_{\text{IC}}} \sum_{i=1}^{N_{\text{IC}}} \left| \hat{x}^{(i)}(t_0) - x_0^{(i)} \right|^2
    \label{eq:LICL}
\end{equation}

Finally, the \emph{physics collocation loss (PCL)} enforces the system dynamics at sampled collocation points. Unlike the physics data loss in \eqref{eq:PLD}, it relies only on network predictions and the governing equations, without requiring ground-truth data, thereby reducing dependence on labeled data. It is defined as:

\begin{equation}
    \mathcal{L}_{\text{PCL}} = \frac{1}{N_c} \sum_{i=1}^{N_c} \left| \frac{d\hat{x}^{(i)}}{dt} - f(t^{(i)}, \hat{x}^{(i)}, u) \right|^2
    \label{eq:PIC}
\end{equation}

In \eqref{eq:LNN}, \eqref{eq:LICL}, and \eqref{eq:PIC}, $N_{\text{d}}$, $N_{\text{IC}}$, and $N_{\text{C}}$ denote the number of labeled data, initial conditions, and collocation points, respectively. The total loss function is defined as:
\begin{align}
    \mathcal{L}_{\mathrm{TOT}} &= \lambda_{\mathrm{NNL}} \mathcal{L}_{\mathrm{NNL}} 
    + \lambda_{\mathrm{PDL}}  \mathcal{L}_{\mathrm{PDL}}
     \notag \\&
    + \lambda_{\mathrm{ICL}}  \mathcal{L}_{\mathrm{ICL}} 
    + \lambda_{\mathrm{PCL}}  \mathcal{L}_{\mathrm{PCL}}
    \label{eq:tot}
\end{align}
where $\lambda_{\mathrm{NNL}}$, $\lambda_{\mathrm{PDL}}$, $\lambda_{\mathrm{ICL}}$, and $\lambda_{\mathrm{PCL}}$ are the respective weighting coefficients. 
\section{Methodology}
\label{sec: Metho}
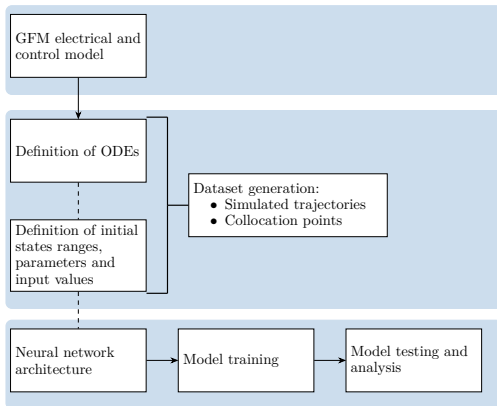
\begin{figure}[H]
    \centering
    \resizebox{0.75\linewidth}{!}{
    \begin{tikzpicture}
        \node(first)[draw, rectangle, text width=3cm, minimum height=1.5cm,fill=white] at (0,0){GFM electrical and control model};
        \node(second)[draw, rectangle, text width=3cm, minimum height=1.5cm, below of =first, node distance=2.5cm,fill=white]{Definition of ODEs};
        \node(third)[draw, rectangle, text width=3cm, minimum height=1.5cm, below of =second, node distance=2.5cm,fill=white]{Definition of initial states ranges, parameters and input values};
        \node(forth)[draw, rectangle, text width=3cm, minimum height=1.5cm, below of =third, node distance=2.5cm,fill=white]{Neural network architecture};
        \node(fifth)[draw, rectangle, text width=3cm, minimum height=1.5cm, right of =forth, node distance=4cm,fill=white]{Model training};
        \node(sixth)[draw, rectangle, text width=3cm, minimum height=1.5cm, right of =fifth, node distance=4cm,fill=white]{Model testing and analysis};
        \node(seventh)[draw, rectangle, text width=4.5cm, minimum height=1.5cm, right of =third, node distance=5cm,yshift=1.2cm,fill=white]{
        Dataset generation:
        \begin{itemize}
            \item Simulated trajectories
            \item Collocation points
        \end{itemize}
        };
        \node[coordinate, left=0.5cm of seventh] (coor) {};
        \node[coordinate, below=0.2cm of first, xshift=10cm] (coorfirst) {};
        \node[coordinate, below=0.2cm of third, xshift=10cm] (coorsecond) {};
        \node[coordinate, below=0.2cm of forth, xshift=10cm] (coorforth) {};
        \draw[myarrow] (first.south)--(second.north);
        \draw[-,dashed] (second.south) -- (third.north);
        \draw[-,dashed] (third.south) -- (forth.north);
        \draw[myarrow] (forth.east) -- (fifth.west);
        \draw[myarrow] (fifth.east) -- (sixth.west);
        \draw[-,thick] (seventh.west) -- (coor.east);
        \draw[-,thick] (coor.north) -- ++(0,2.1cm) coordinate (coor2);
        \draw[-,thick] (coor2.west) -- ++(-0.5cm,0);
        \draw[-,thick] (coor.south) -- ++(0,-2.1cm) coordinate (coor3);
        \draw[-,thick] (coor3.west) -- ++(-0.5cm,0);
        \begin{pgfonlayer}{background}
			\node[fit=(first.north) (first.west) (coorfirst) (first.south), inner xsep=0.1cm, inner ysep=0.2cm, 
			fill=CustomBlue!20, rounded corners,
			draw=CustomBlue!20, thick] {};
		\end{pgfonlayer}
        \begin{pgfonlayer}{background}
			\node[fit=(second.north) (second.west) (coorsecond) (third.south), inner xsep=0.1cm, inner ysep=0.2cm, 
			fill=CustomBlue!20, rounded corners,
			draw=CustomBlue!20, thick] {};
		\end{pgfonlayer}
        \begin{pgfonlayer}{background}
			\node[fit=(forth.north) (forth.west) (coorforth) (forth.south), inner xsep=0.1cm, inner ysep=0.2cm, 
			fill=CustomBlue!20, rounded corners,
			draw=CustomBlue!20, thick] {};
		\end{pgfonlayer}
    \end{tikzpicture}}
    \caption{Overall methodology followed in this paper}
    \label{fig:Metho}
\end{figure}

This section presents the methodology adopted to train a PINN capable of capturing the dynamics of the GFM. The overall workflow is illustrated in Figure \ref{fig:Metho}.
\subsection{Dataset Generation}
The ODEs of the GFM are first derived as described in Section~\ref{subsec:model}, and stable operating ranges are defined for the system states. Trajectories are generated using an RK45 solver with initial conditions sampled via Latin hypercube sampling (LHS). The reference voltage and frequency in \eqref{eq: inputs} are kept constant at $20~\mathrm{kV}$ and $50~\mathrm{Hz}$, respectively, while the remaining reference values are randomly sampled within predefined ranges. The grid parameters and control gains are fixed based on \cite{winkens2021impact}. Labeled input–output trajectories and collocation points are generated to enforce physics-based constraints. The PINN is compared with a VNN, where both are trained on the same 500 labeled trajectories, while the PINN additionally uses 16{,}000 sampled initial conditions for collocation points. 

\subsection{Physics-Informed Neural Network Training}
The generated trajectories were divided into $80\%$ training, $10\%$ validation, and $10\%$ test sets. The neural network was implemented in \texttt{PyTorch} using the toolbox from~\cite{karampinis2025pinnstoolbox} and consisted of four hidden layers with 128 neurons each using the hyperbolic tangent (tanh) activation function. Training was performed with the \texttt{ADAM} optimizer using a learning rate of 0.001 for 1200 epochs and the smooth loss criterion. The loss weights were set to $\lambda_{\mathrm{NNL}}=1$, $\lambda_{\mathrm{PDL}}=10^{-5}$, $\lambda_{\mathrm{ICL}}=0.1$, and $\lambda_{\mathrm{PCL}}=10^{-5}$. Initial weight values were taken from \cite{karampinis2025pinnstoolbox} and further refined through multiple training runs to identify the configuration yielding the lowest loss. Hyperparameters were manually tuned through extensive experiments, while future work will investigate further optimization to improve model performance. Weights \& Biases was used for monitoring and hyperparameter tuning. Training was carried out on the RWTH Aachen HPC cluster using a 96-core Intel Xeon Platinum 8468 CPU node with 128 GB RAM and an NVIDIA H100 GPU.

\subsection{Case Studies and Analysis}
We conducted two main case studies:

\begin{itemize}
\item We compared the PINN and VNN in terms of prediction accuracy using root mean-square error (RMSE), mean absolute error (MAE), and maximum AE.
\item We compared the inference time of the PINN, RK45 solver, and VNN.

\end{itemize}

\section{Results and Analysis}
\label{sec: Results}
\subsection{PINN Predictions}
The final training, validation, and test losses for both the PINN and VNN are shown in Table~\ref{tab: loss}. The results clearly indicate improved performance for the PINN, which achieves lower training and test losses on the same dataset, demonstrating its superior ability to learn the system dynamics and produce more accurate predictions.

\begin{figure*}[h!]
    \centering
    \resizebox{0.98\linewidth}{!}{
    \includegraphics[width=\linewidth]{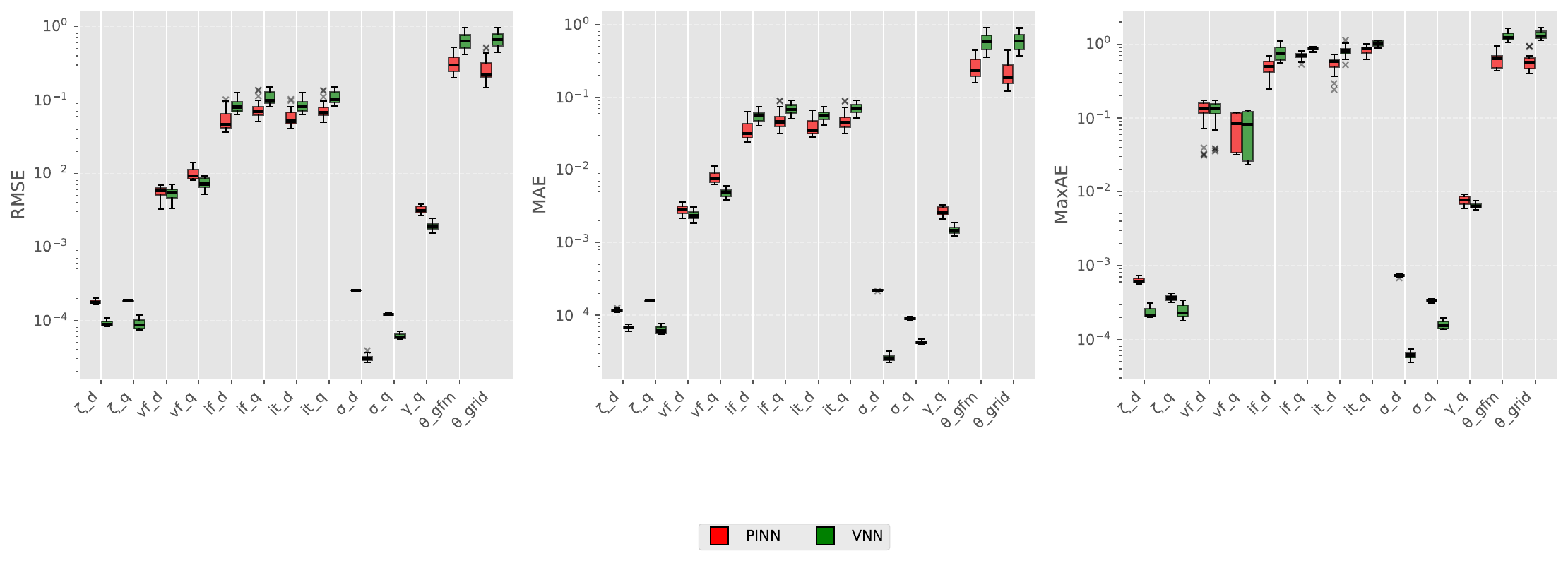}}
    \caption{RMSE, MAE, and MaxAE metrics computed for each state on the test dataset for both the PINN and the VNN}
    \label{fig:boxplot}
\end{figure*}

\begin{table}[H]
\centering
\caption{Comparison of training, validation, and test losses for PINN and VNN}
\label{tab: loss}
\begin{tabular}{lccc}
\hline
\textbf{Model} & \textbf{Training Loss} & \textbf{Validation Loss} & \textbf{Test Loss} \\
\hline
PINN & 0.0036 & 0.0024 & 0.0085 \\
VNN  & 0.0181 & 0.0059 & 0.0378 \\
\hline
\end{tabular}
\end{table}
Figure~\ref{fig:boxplot} presents boxplots of prediction errors over 20 test trajectories with 1200 samples each, highlighting the superior consistency and robustness of the PINN compared to the VNN. The PINN achieves lower median errors and smaller interquartile ranges for current states, indicating higher accuracy and stability, while both models show similar performance for voltage states. Although the VNN performs slightly better for some voltage-related outputs, the improvement is minor and more variable. The MaxAE results further demonstrate the PINN’s reduced extreme errors for currents. Overall, the physics-informed formulation improves prediction accuracy and reliability, which is important for real-time dynamic stability assessment. Both models required similar training times of approximately 12 hours.
  
\begin{figure}[H]
    \centering

    \begin{subfigure}{0.49\textwidth}
        \centering
        \includegraphics[width=\linewidth]{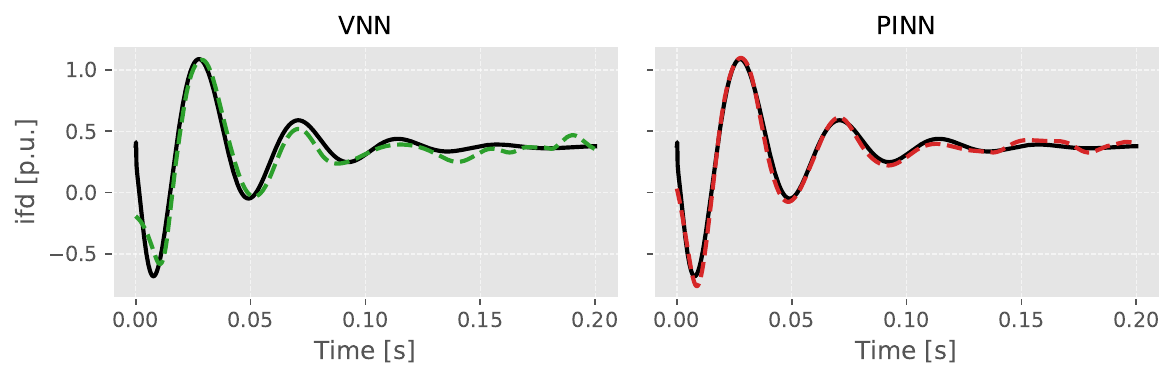}
        \label{fig:ifd}
    \end{subfigure}
    \hspace{0.001\textwidth}
    \begin{subfigure}{0.49\textwidth}
        \centering
        \includegraphics[width=\linewidth]{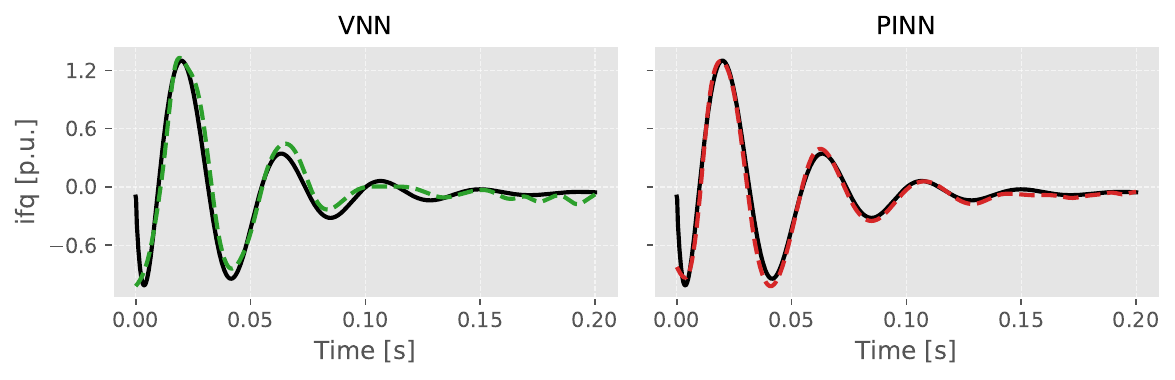}
        \label{fig:ifq}
    \end{subfigure}

    \caption{$dq$-components of the filter current with ground truth in black and dashed predictions}
    \label{ifd_q}
\end{figure}

\begin{figure}[H]
    \centering

    \begin{subfigure}{0.49\textwidth}
        \centering
        \includegraphics[width=\linewidth]{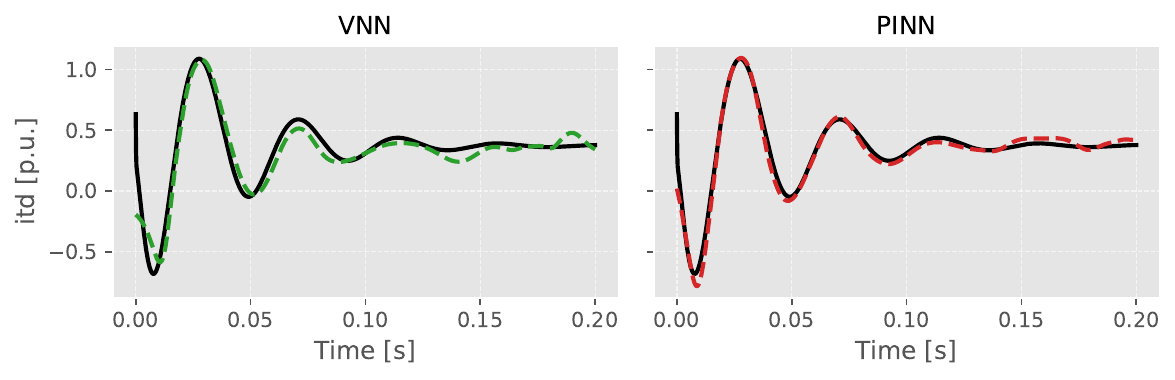}
        
    \end{subfigure}
    \hspace{0.001\textwidth}
    \begin{subfigure}{0.49\textwidth}
        \centering
        \includegraphics[width=\linewidth]{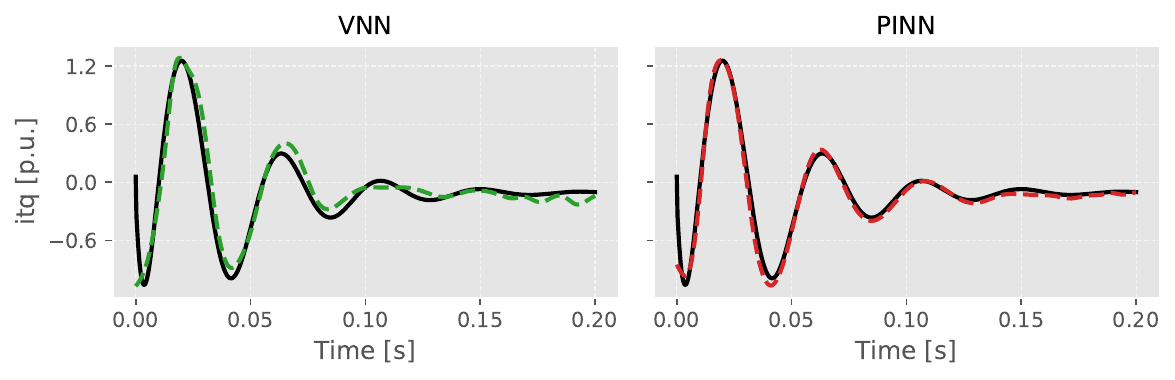}
       
    \end{subfigure}
    \caption{$dq$-components of the grid-side current}
    \label{itdq}
\end{figure}

\begin{figure}[H]
    \centering

    \begin{subfigure}{0.49\textwidth}
        \centering
        \includegraphics[width=\linewidth]{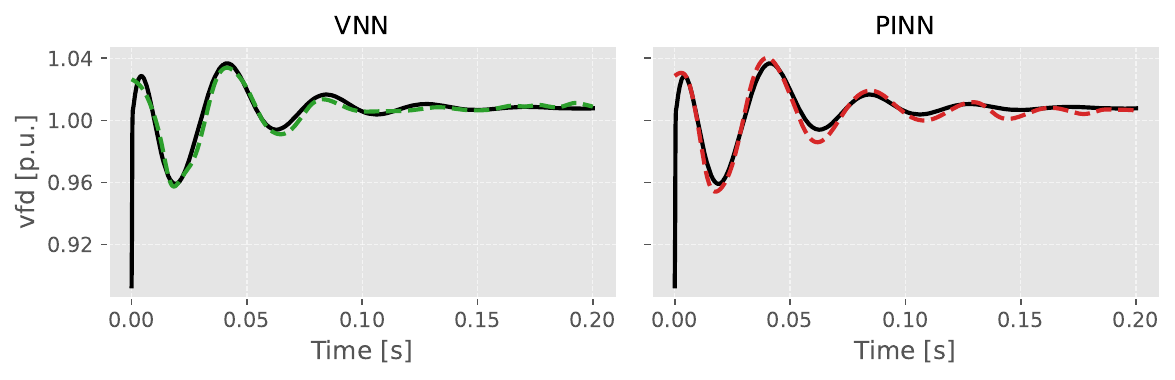}
        
    \end{subfigure}
    \hspace{0.001\textwidth}
    \begin{subfigure}{0.49\textwidth}
        \centering
        \includegraphics[width=\linewidth]{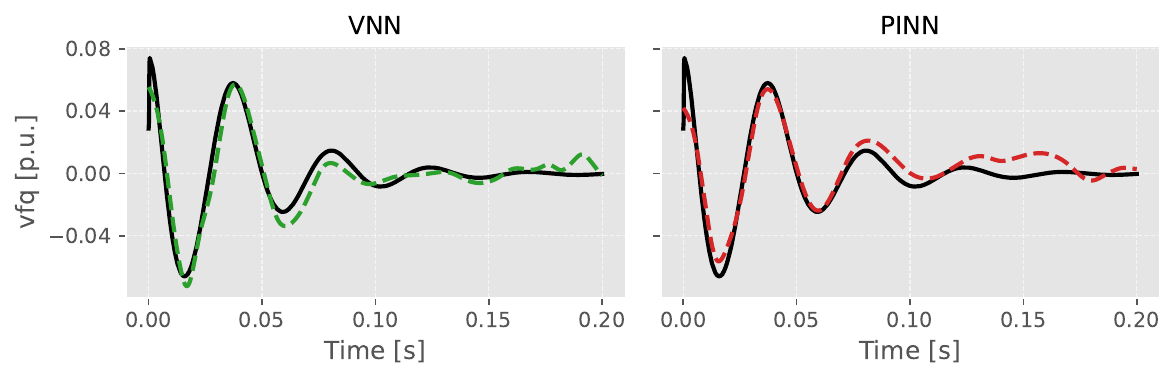}
        
    \end{subfigure}
    \caption{$dq$-components of the filter voltage}
    \label{vfdq}
\end{figure}

To demonstrate the ability of the trained PINN to reproduce grid-forming converter dynamics, Figures~\ref{ifd_q}, \ref{itdq}, and \ref{vfdq} show the predicted voltage and current trajectories for both the VNN and the PINN. While the VNN performs slightly better for the voltage states, the PINN provides significantly more accurate current predictions. Neither model, however, accurately predicts the initial conditions, likely due to the steep and discontinuous behavior immediately after the initial instant.

\subsection{Inference time}
\begin{figure}[H]
    \centering
    \includegraphics[width=0.8\linewidth]{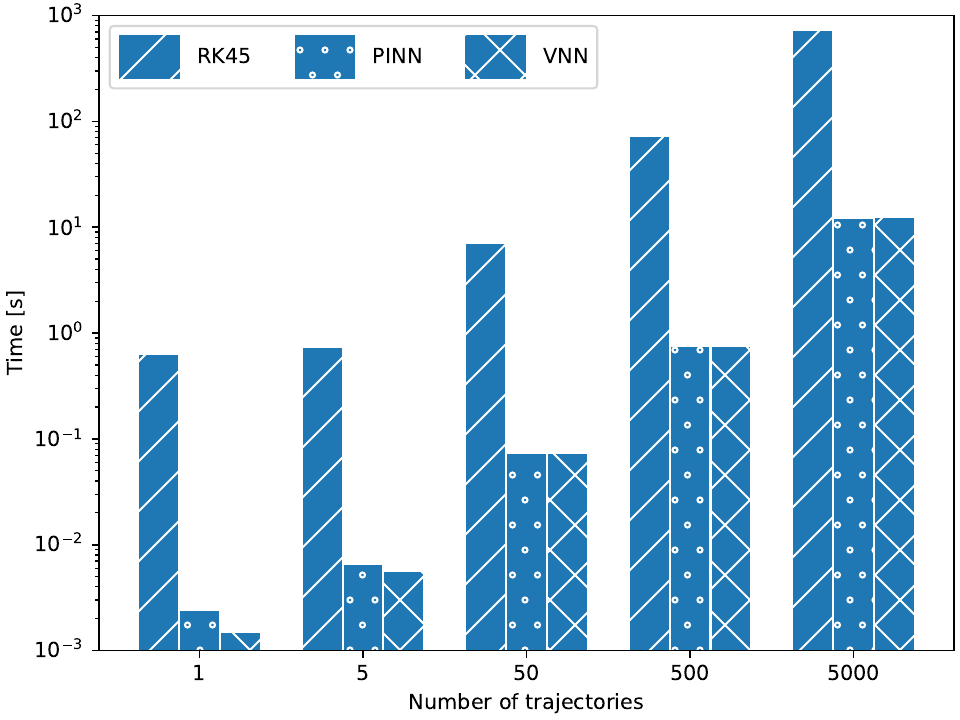}
    \caption{Inference time for different number of trajectories}
    \label{fig:placeholder}
\end{figure}

Figure~\ref{fig:placeholder} compares the inference time of RK45, the PINN, and the VNN for different trajectory counts. The PINN is much faster than RK45, reaching over 400× speedup for a single trajectory and nearly 80× for larger ones, showing strong potential for real-time applications. Unlike RK45, whose computation time grows almost linearly due to step-by-step integration, the PINN and VNN generate trajectories in a single forward pass using matrix multiplications.
\section{Discussion and Conclusion}
\label{sec: Conc}
In this paper, we propose a PINN to model the full dynamic behavior of a droop-controlled GFM. The model incorporates the governing physical equations into the loss function, improving prediction accuracy, reducing training data requirements, and ensuring physically consistent results. We also compare the inference time of the proposed network with traditional numerical solvers.

The experimental results show that the PINN consistently outperforms VNN when both are trained and tested using the same amount of labeled data. For current prediction, the PINN clearly surpasses the VNN, while for voltage prediction, the VNN performs slightly better. In terms of computational speed, both the PINN and the VNN demonstrate substantially faster inference times compared to RK45. This highlights the potential of neural-network-based surrogate models for real-time DSA. Given the promising performance of PINNs in predicting converter dynamics, a natural extension of this work is to apply the method to other converter types, such as GFL. Furthermore, we plan to integrate PINN models in simulations. 



\bibliography{ifacconf}             
                                                   







\appendix
\end{document}